\documentclass[aps,prx,twocolumn,superscriptaddress]{revtex4-2}

\usepackage{amsmath,amssymb}
\usepackage[pdftex]{hyperref,graphicx}
\hypersetup{colorlinks = true, urlcolor = blue, linkcolor = blue, citecolor = blue}

\usepackage{physics}
\usepackage{xcolor}
\usepackage{bm}

\begin{document}

	\title{Disorder-induced zero-bias peaks in Majorana nanowires}

	\author{Sankar Das Sarma}
	\affiliation{Condensed Matter Theory Center and Joint Quantum Institute, Department of Physics, University of Maryland, College Park, Maryland 20742, USA}

	\author{Haining Pan}
	\affiliation{Condensed Matter Theory Center and Joint Quantum Institute, Department of Physics, University of Maryland, College Park, Maryland 20742, USA}
	\affiliation{Kavli Institute for Theoretical Physics, University of California, Santa Barbara, CA 93106, USA}
	
\begin{abstract}
	Focusing specifically on the recently retracted work by Zhang \textit{et al.} [H. Zhang \textit{et al}., \href{https://doi.org/10.1038/nature26142}{Nature (London) \textbf{556}, 74 (2018)}; Retraction, \href{https://www.nature.com/articles/s41586-021-03373-x}{Nature (London) \textbf{591}, E30 (2021)}] and the related recently available correctly analyzed data from this Delft experiment [H. Zhang \textit{et al}., \href{http://arxiv.org/abs/2101.11456}{arXiv:2101.11456}], we discuss the general problem of confirmation bias in experiments verifying various theoretical topological quantization predictions.  We show that the Delft Majorana experiment is most likely dominated by disorder, which produces trivial (but quite sharp and large) zero-bias Andreev tunneling peaks with large conductance $ \sim 2e^2/h $ in the theory, closely mimicking the data. Thus, although the corrected Delft data are by far the best tunnel spectroscopy results available in the literature, manifesting large and sharp zero-bias peaks rising above the background with an impressive hard superconducting gap, our theory shows that the most natural explanation for these zero-bias peaks is that they are disorder-induced and not topological Majorana modes. It is possible to misinterpret such disorder-induced zero-bias trivial peaks as the apparent Majorana quantization, as was originally done in 2018 arising from confirmation bias.  One characteristic of the disorder-induced trivial peaks is that they manifest little stability as a function of Zeeman field, chemical potential, and tunnel barrier, distinguishing their trivial behavior from the expected topological robustness of non-Abelian Majorana zero modes. We also analyze a more recent nanowire experiment [Yu \textit{et al}., \href{https://www.nature.com/articles/s41567-020-01107-w}{Nat. Phys. \textbf{17}, 482 (2021)}] which is known to have a huge amount of disorder, showing that such highly disordered nanowires may produce very small above-background trivial peaks with conductance values $ \sim 2e^2/h $ in a dirty system manifesting very soft superconducting gap with substantial in-gap conduction, as were already reported by several groups almost 10 years ago. Removing disorder and producing cleaner samples through materials quality improvement and better fabrication is the only way for future progress in this field.
\end{abstract}
\maketitle

\section{Background and Introduction}\label{sec:I}

Topological phenomena have been among the most active research areas in condensed matter physics during the last 40 years ever since the discovery of quantum Hall effects~\cite{vonklitzing1980new,tsui1982twodimensional} in the early 1980s.  It is now well accepted that the precise quantization of Hall conductance in units of $ e^2/h $ is a topological effect, thus explaining both its exactness and its robustness. In fact, the Hall quantization is so precise in the quantum Hall effect, that this has now become the official resistance standard defining the ``Ohm", thus topological quantization has now transcended basic science, becoming a part of everyday usage in electrical appliances everywhere. The quantization is protected by a bulk gap (``topological gap'') along with gapless (``zero-energy'') boundary zero modes.  The reason integer quantum Hall effect~\cite{vonklitzing1980new} is more robust than the fractional one~\cite{tsui1982twodimensional} is simply that the gap is much larger in the integer case. The larger the gap, the more robust and precise the quantization. This topological theoretical understanding of quantum Hall effects was developed after the experimental discoveries, not before, and the quantum Hall effect was not a theoretical prediction, it was an experimentally discovered phenomenon during 1980--1982.  

By contrast, topological theoretical predictions have dominated condensed matter physics during the last 20 years with the worldwide extensive experimental search looking for specific theoretical predictions of quantization effects.  Among the more well known of such predictions being studied experimentally are {quantum spin Hall effect}~\cite{konig2007quantum}, {topological insulator}~\cite{chen2009experimental}, {quantum anomalous Hall effect}~\cite{chang2013experimental}, {non-Abelian braiding}~\cite{willett2009measurement}, {Kitaev spin liquid}~\cite{kasahara2018majorana}, {Majorana zero mode}~\cite{mourik2012signatures,zhang2018quantizeda}, etc.  Many of these theoretical predictions are precise, for example, quantum spin Hall effect, quantum anomalous Hall effect, and Majorana zero mode are associated with conductance quantization, protected by an energy gap in the topological phase. The topological insulator and Kitaev spin liquid come with theoretical predictions of quantization in the electromagnetic and thermal response, respectively.  This raises the serious problem of potential confirmation bias in the putative topological experimental discoveries often claimed in the literature since the theoretical prediction is precise, and condensed matter physics imposes no community standards on the definition of an experimental discovery as is common in high-energy physics.  This is particularly problematic for topological discoveries since the samples are often rather complex with multiple different materials layers (which are always unintentionally disordered), and topology often applies only when the sample size (temperature or dissipation or disorder broadening) exceeds (is below) some characteristic topological length (the topological gap), giving the experimentalist considerable leeway in generously explaining away deviations in the experimental ‘quantization’ from the precise theoretical value as arising from finite size and/or small energy gap and/or finite broadening/temperature effects.  

A problem of considerable significance, which requires close scrutiny, is the key role confirmation bias may play in the experimental claims of the various topological discoveries, which have been precisely theoretically predicted.  When one knows exactly what one is looking for among the huge amount of the collected data, typically as a function {of} many experimental tuning parameters in many samples, and one has considerable freedom in what one may report in the experimental publications, fine tuning, data selection, and even confirmation bias may creep in unwittingly and unintentionally in verifying the known theoretical prediction.  In fact, the somewhat unsatisfactory situation existing in the quantum spin Hall~\cite{lunczer2019approaching} conductance quantization {has} recently been discussed in depth, and a recent {claim}~\cite{he2017chiral} for Majorana modes in the quantum anomalous Hall scenario has been forcefully debunked~\cite{kayyalha2020absence}.   Similarly, experimental claims of Majorana zero modes in ferromagnetic {chains}~\cite{nadj-perge2014observation} were persuasively shown to be likely innocuous signatures of nontopological Shiba states in a rather complicated situation~\cite{sau2015bound,ruby2015end,wang2021spinpolarized}.  Similarly, a compelling claim for the observation of non-Abelian quasiparticle interference in the $ \frac{5}{2} $  fractional quantum Hall state has never been reproduced and is generally ignored in the literature in spite of the singular importance of the claim itself~\cite{willett2009measurement}.    Another important {claim}~\cite{albrecht2016exponential} for the observation of exponential protection of topological Majorana modes in nanowires has been shown to be arising from trivial modes with the claimed exponential wire length dependence being an artifact of having too few {samples}~\cite{chiu2017conductance,lai2021theory}. One can actually provide many more such examples of topological claims, which are not reproduced by others and are generally ignored subsequently, e.g., the observed thermal quantization in a Kitaev spin liquid~\cite{kasahara2018majorana}.  

Another class of topological discovery claims, precisely following theoretical predictions, involves the so-called three-dimensional (3D) topological Dirac and Weyl materials, which have linearly dispersing semimetallic band structures with certain ``topological" properties induced by the multi-valley structure of the system along with spin-valley-orbital coupling.  For example, the claimed appearance of {double Fermi arcs}~\cite{xu2015discovery,xu2015discoverya,lv2015observation,neupane2014observation} in certain situations was shown to arise from well-understood nontopological physics~\cite{kargarian2016are}.  Various sensational claims of the observation of the axial anomalies in Dirac-Weyl systems based just on the mere observation of negative magnetoresistance have been shown to be misleading as the observed effect most likely arises from ionic impurity scattering in a magnetic field~\cite{pixley2016disorderdriven}.   There are far too many examples of such topological discoveries arising from fine-tuned (and often carefully post-selected) data precisely following theoretical predictions.

In all of these situations, the experimental claims of possible topological discovery were published with much fanfare only because there were compelling earlier theoretical predictions; the experimental data by itself in none of these cases is so spectacular (i.e., without the theoretical predictions) as to merit great attention.  In such situations, attention to and zealously guarding against data selection and confirmation bias is essential. In particular, it becomes imperative to trenchantly analyze the claimed topological discovery, not only through extensive independent data analyses, but also through rigorous theoretical scrutiny, to ensure that possible nontopological (or other trivial) explanations can be decisively ruled out.  Also, a general community standard should be to await faithful experimental reproduction of the claim by independent experimental groups before the acceptance of the claim, no matter how compelling the agreement between the original experimental claim and the theoretical prediction might be. 

In this paper, we provide a rather compelling theoretical analysis with concrete simulations for one of the most high-impact recent topological discovery claims (on Majorana zero modes) from Delft~\cite{zhang2018quantizeda}, which has just been retracted~\cite{zhang2021retraction} as the authors concluded on the reanalysis and recalibration of the original data that the original claim might have suffered from confirmation bias with fine-tuned data selection unwittingly preferring false positives in favor of the known theoretical prediction~\cite{sau2010nonabelian}.   This retraction~\cite{zhang2021retraction} of the high-profile article~\cite{zhang2018quantizeda} in favor of a new detailed article~\cite{zhang2021large} with corrected data and analysis has led to the conclusion that the original claim of the Majorana quantization observation is untenable.  Although this is the only retraction of a high-profile topological experimental discovery claim in the literature, we believe that similar confirmation bias (but not necessarily the specific data problems arising in~\cite{zhang2021retraction,zhang2018quantizeda})  applies to many other topological discovery claims in the literature during 2000--2020  where a precise knowledge of what one is looking for has been the key factor in the discovery claim, with the experimental quantization results themselves not being sufficiently compelling without the existing knowledge of the theoretical prediction coupled with extensive fine tuning and postselection of data ensuring it fits the known theoretical prediction.  

Since the corrected Delft results presented in~\cite{zhang2021large} are by far the best tunneling conductance data connected with Majorana physics in the literature, because of the very large and sharp zero-bias conductance peak values $ \sim2e^2/h $ reported in this work, it is imperative that we develop a theoretical analysis of the corrected Delft data to understand the underlying physics.  Although our theory and analysis in this work are specific only to the Majorana nanowire physics of~\cite{zhang2021large}, we believe that our conclusions are far reaching and may have substantial relevance to the somewhat unsatisfactory aspect of various experimental claims in the whole field of topological condensed matter physics.  Our results certainly apply to most of the Majorana experiments during 2012--2021 in the literature, which were trying to verify precise theoretical predictions made in 2010~\cite{sau2010nonabelian,sau2010generic,lutchyn2010majorana,oreg2010helical,kitaev2001unpaired} claiming evidence in support of the existence of Majorana zero modes in one-dimensional (1D) nanowires. We also provide a brief analysis of a very recent nanowire experiment with small peaks, which most obviously arises from disorder~\cite{yu2021nonmajorana}. The data presented in this recent experiment~\cite{yu2021nonmajorana} are taken on samples with very high amount of disorder, with substantial subgap conduction and a very soft gap, but still manifesting small conductance peaks which can occasionally be fine tuned to values of $ O(2e^2/h) $. Such fine tuning and postselection merely to fit the data to the known prediction is extremely dangerous, unless done with great care and rigor, as the retraction of Zhang~\cite{zhang2018quantizeda} demonstrates vividly~\cite{zhang2021retraction}. 

{We emphasize that we do not claim, by any stretch of the imagination, that all reported experimental topological discoveries of the last 20 years are suffering from confirmation bias and are consequently suspect.  We are pointing out the confirmation bias issue as a genuine problem that one must always take into account in analyzing experimental results and forming a conclusion if the experiment is purported to verify well-defined theoretical topological predictions.  In particular, the observation of an approximate quantization in only a small fine tuned fraction of the data, which has been postselected, should not either be construed as the discovery of the quantization without carefully considering alternate possibilities.  Reproducing the quantization in multiple samples without excessive postselection and fine tuning should be a standard part of the measurement protocol if the measured data are verifying precise theoretical predictions.  All the experimental data should be made available to the whole community in claiming topological quantization of any kind so that there can be independent and objective checks that confirmation bias did not affect the results in an unintended manner.  Our goal in this work is to provide a specific example of such a confirmation bias playing a detrimental role in one of the most well-known topological discoveries of the last 20 years, namely, the quantized Majorana conductance in semiconductor nanowires~\cite{zhang2018quantizeda}, and not to impugn that many topological discoveries might have been definitely compromised by confirmation bias.  We are alerting the community about the possibility of confirmation bias, using a concrete example, when complex measurements on highly complicated samples are carried out in order to verify established theoretical predictions.}

In Sec.~\ref{sec:II}, we present our theory and describe our results and simulations.  We conclude in Sec.~\ref{sec:III}, discussing the implications of our theoretical findings to the field of Majorana zero modes specifically, and topological condensed matter physics generally. Our work provides a clear theoretical interpretation for the existing nanowire ``Majorana'' data, and compellingly demonstrates the serious problem of achieving fine-tuned experimental agreement with theoretical predictions as new theoretical understanding develops, making the fine-tuned data selection look like dubious confirmation bias.

\section{Theory and Results}\label{sec:II}
\begin{figure}[t]
	\centering
	\includegraphics[width=3.4in]{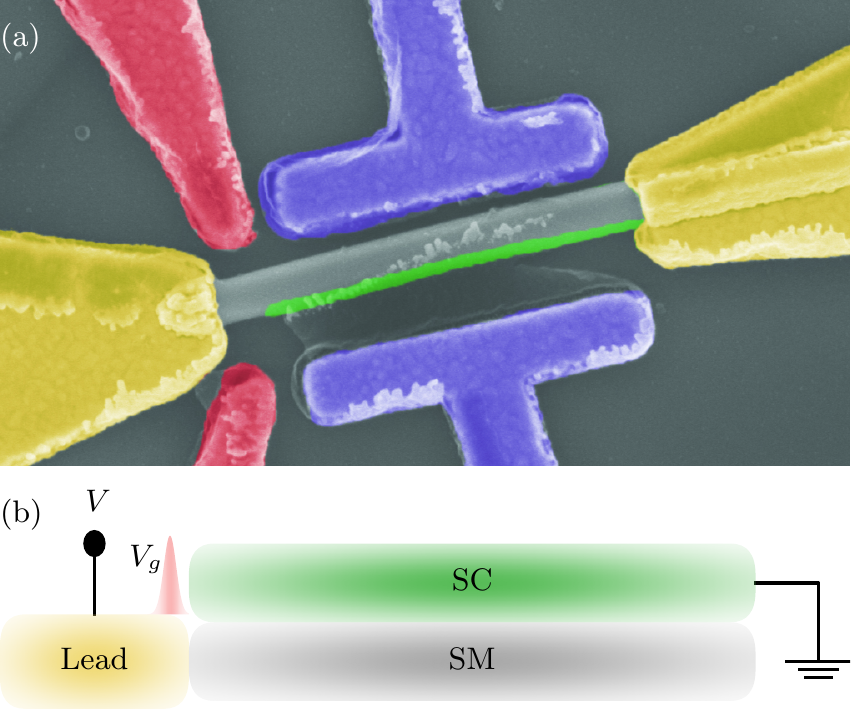}
	\caption{(a) The experimental device from Fig. 1 of Ref.~\onlinecite{zhang2021large}. The InSb nanowire (gray) is covered by the Al shell (green). The tunnel gate (red) controls the tunnel barrier $ V_g $ in (b). The super gate (purple) controls the chemical potential. The normal lead (yellow) is attached to one end of the nanowire.
	(b) The schematic of the semiconductor-superconductor hybrid nanowire in the theory. The bias voltage is applied to the lead and the superconductor is grounded.}
	\label{fig:1}
\end{figure}

We show in Fig.~\ref{fig:1} the experimental structure [reproduced directly from Fig. 1(a) of~\cite{zhang2021large}] as well as the corresponding 1D theoretical idealization studied in this work.  A semiconductor (InSb) nanowire is in contact with a superconductor (Al) so that the nanowire becomes superconducting due to the proximity effect~\cite{doh2005tunable}.  The experiment (and our theory) studies tunneling spectroscopy through the nanowire using the standard normal-superconductor (NS) tunneling structure at one end (left end in Fig.~\ref{fig:1}) by controlling a tunnel barrier (seen in red as tunnel gates in Fig.~\ref{fig:1}) in the presence of a magnetic field applied parallel to the nanowire.  In addition to superconductivity and field-induced Zeeman spin splitting ($ V_z $), the third important physics ingredient is the spin-orbit coupling in the nanowire.  It was predicted in Refs.~\onlinecite{sau2010nonabelian,lutchyn2010majorana,oreg2010helical,kitaev2001unpaired,sau2010generic}  that such a 1D wire, in the presence of superconductivity and spin-orbit coupling would lead to a field-tuned topological quantum phase transition (TQPT) at a critical Zeeman splitting $ V_c = (\Delta^2 + \mu^2) ^{1/2} $, where $ \Delta $ and $ \mu $ are, respectively, the induced superconducting gap and the chemical potential in the nanowire, with $ V_z > V_c (V_z < V_c) $ being the topological (trivial) regimes, respectively.  Majorana zero mode (MZM) appears at zero energy at both ends of the wire for $ V_z > V_c $ as the system enters the topological regime with a topological gap opening up; the gap vanishes at the TQPT by definition.
Our theory is the standard free-fermion Bogoliubov-de Gennes(BdG) theory for the nanowire structure shown in Fig.~\ref{fig:1} including, in addition to the superconducting pairing term, a spin-orbit coupling, a Zeeman field, a potential disorder, and a self-energy term arising from integrating out the degrees of freedom associated with the parent superconductor Al~\cite{pan2020physical}.  The BdG equation is solved numerically exactly, and then the resultant eigenstates (i.e., wave functions) and eigenvalues (i.e., energies) are used to construct a scattering matrix to solve the NS tunneling problem exactly, using the KWANT $S$-matrix code~\cite{groth2014kwant},  to calculate the tunneling spectrum obtaining the conductance as a function of bias voltage, Zeeman field, chemical potential, and the tunnel barrier controlling the tunneling amplitude at the NS junction.  The theory should be thought of as a first-principles conductance transport calculation using an exact numerical ``band structure" as the input.  In such a first-principles transport theory, the important points are the results, not the theory itself, which is standard and well established.  We, therefore, refer to our earlier work~\cite{pan2020physical}  for the theoretical details.  We use the same parameters for the InSb/Al system as in Ref.~\onlinecite{pan2020physical}, and we provide these parameters in the caption of our Fig.~\ref{fig:1}.  The parameters correspond to the InSb/Al structures used in Ref.~\onlinecite{zhang2021large} which we are trying to understand.

The most important aspect of the theory, which in our opinion is also the key physics underlying the zero-bias conductance peaks reported in Majorana experiments, is the inclusion of a model spatially random potential, $ V(x) $, added to the constant chemical potential $ \mu $. This random term represents the disorder invariably present in the nanowire due to the presence of quenched random impurities and interface imperfections (and perhaps elsewhere, e.g., the substrate and tunnel junction).  The presence of $ V(x) $ in the BdG Hamiltonian makes our theory correspond to the ``ugly'' situation in the terminology of Ref.~\onlinecite{pan2020physical},  whereas for $ V(x)=0 $ the system is pristine, corresponding to the ``good'' situation, which is the situation the experimentalists always have in their mind as the theoretical prediction to be emulated.  The good situation is the standard theoretical Majorana scenario~\cite{sau2010nonabelian,sau2010generic,lutchyn2010majorana,oreg2010helical,kitaev2001unpaired}, where topological zero-bias conductance peaks (ZBCPs) appear in the tunneling spectra for $ V_z > V_c $, whereas for $ V_z < V_c $ the subgap conductance is basically zero.  The presence of random disorder is the key physical mechanism in our theory which, in our opinion, produces much of the physics showing up in the experimental ZBCPs reported in Refs.~\onlinecite{zhang2018quantizeda,zhang2021retraction,zhang2021large,yu2021nonmajorana}.  We contend that the physics of Majorana nanowires at this point of time is dominated by disorder rather than topology as discussed below. Although there has been impressive materials development  (e.g., hard gap~\cite{chang2015hard}) and experimental tunnel conductance has evolved during 2012--2021, we contend that all reported Majorana nanowire data are likely to be disorder dominated and do not reflect the pristine ``good'' predictions for topological Majorana zero modes.
\begin{figure*}[t]
	\centering
	\includegraphics[width=6.8in]{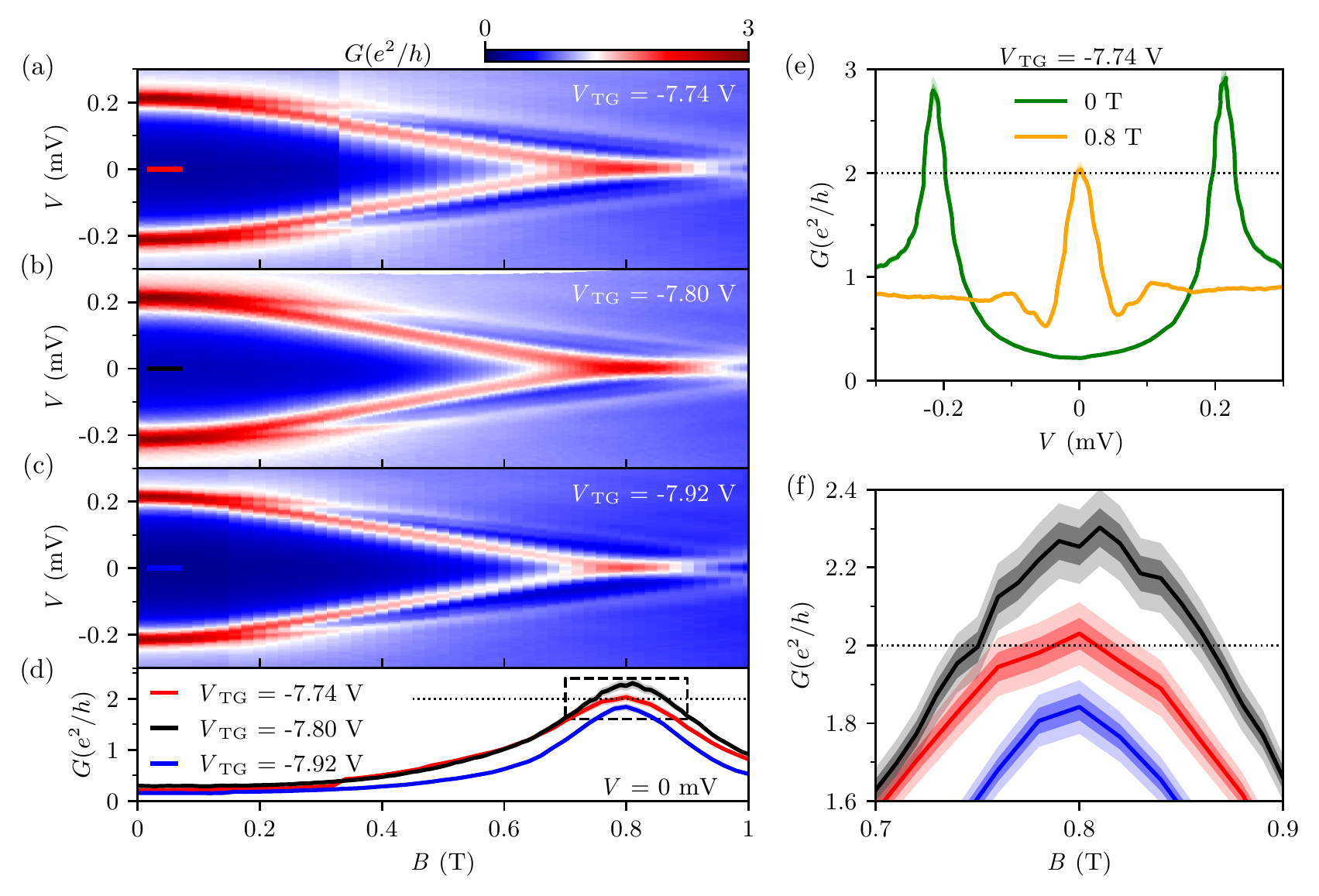}
	\caption{The experimental tunneling conductances from Ref.~\onlinecite{zhang2021large}. 
	Panels (a)-(c) are tunneling conductances as a function of the bias voltage and magnetic field at various tunnel gate voltages $ V_{\text{TG}}$=$ -7.74 $, $ -7.80 $, $ -7.92 $ V, respectively. 
	(d) The horizontal line cuts at zero bias as a function of magnetic field corresponding to (a)-(c).
	(e) The vertical line cuts at zero magnetic field (red) and finite magnetic field (orange) at $ V_{\text{TG}}=-7.74 $ V.
	(f) The closed-up view of the conductance peak near $ 2e^2/h $, indicated by the dashed rectangle in (d). The shading regions indicate the error bar of $ 1\sigma $ (darker) and $ 2\sigma $ (lighter).
	Refer to Ref.~\onlinecite{zhang2021large} for the other parameters of gate voltages.}
	\label{fig:2}
\end{figure*}

In our Fig.~\ref{fig:2}, we reproduce Fig. 2 from Ref.~\onlinecite{zhang2021large} exactly as it appears in this experimental work with no modification because these results shown in Fig.~\ref{fig:2} are what we are trying to explain theoretically.  In Fig.~\ref{fig:2}, what are shown are the measured conductance spectra as a function of the magnetic field, bias voltage, and tunnel barrier.  The theory must be able to reproduce all three parameter dependence faithfully for us to claim understanding.  The experimental measurement temperature being very low ($ \sim $ 20 mK), we show our theoretical results at $ T=0 $ with no loss of generality. Thermal effects are easy to include and would not modify the results as long as the thermal broadening is smaller than the tunneling energy, which is easily checked in the experiment by ensuring that the ZBCP saturates with the lowering of temperature.

\begin{figure*}[t]
	\centering
	\includegraphics[width=6.8in]{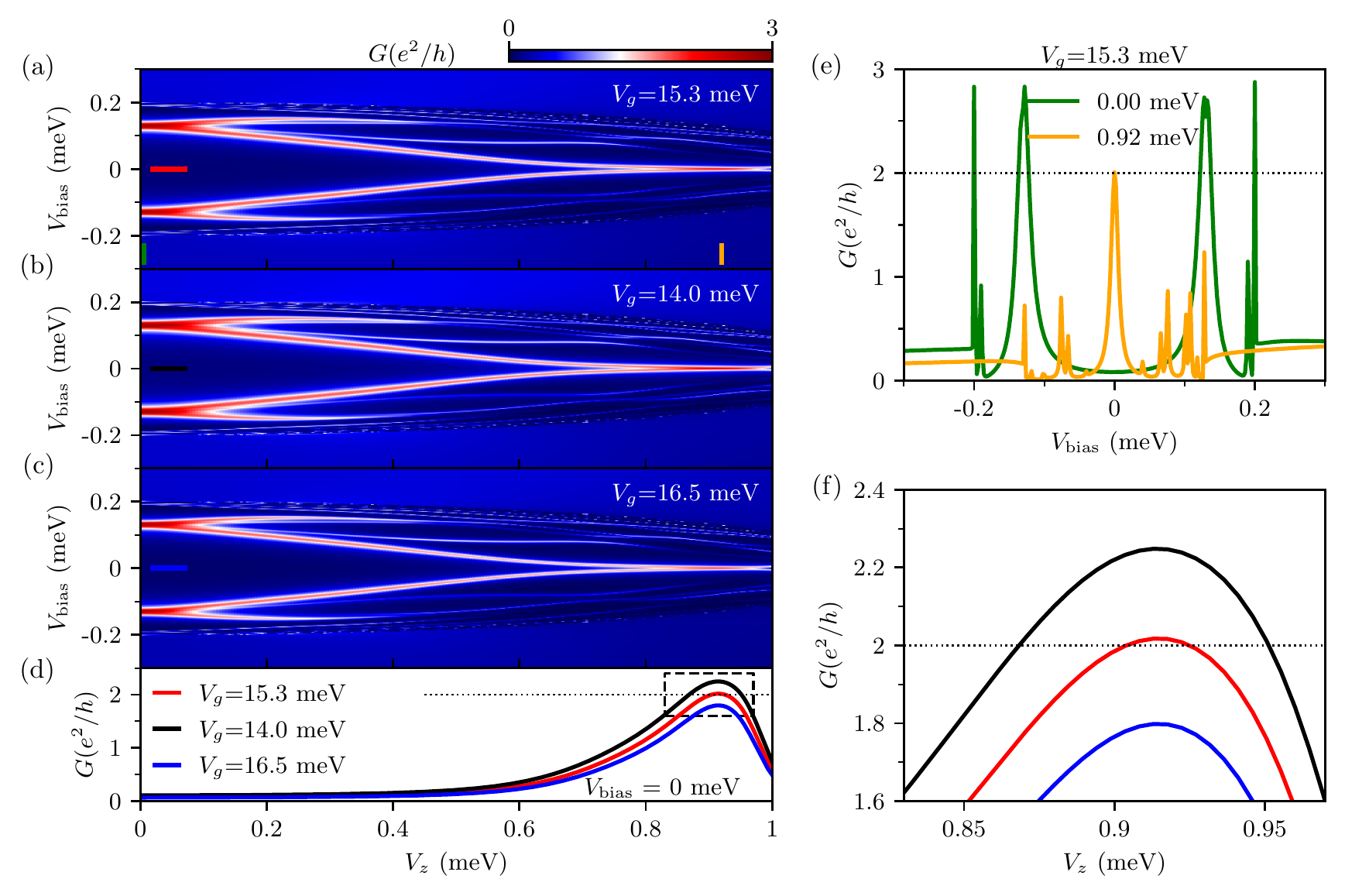}
	\caption{The theoretical tunneling conductances in the trivial regime which qualitatively reproduce the experimental measurements in Fig.~\ref{fig:2}. 
	Panels (a)-(c) are tunneling conductances as a function of the bias voltage and Zeeman field at different tunnel barriers $ V_g$=15.3, 14.0, and 16.5 meV, respectively.
	(d) The horizontal line cuts at zero bias as a function of Zeeman field corresponding to (a)-(c).
	(e) The vertical line cuts at zero magnetic field (red) and finite magnetic field (orange) at $ V_g=15.3 $ meV.
	(f) The closed-up view of the conductance peak near $ 2e^2/h $. Note that the error bar is not applicable here compared to Fig.~\ref{fig:2} because it is purely the theoretical calculation. The parameters are as follows: the wire length is 1 $\mu $m, the chemical potential $ \mu=1 $ meV, the variance of disorder $ \sigma_\mu=1 $ meV, the parent SC gap $ \Delta_0=0.2 $ meV~\cite{lutchyn2018majorana}, the SC-SM coupling strength $ \gamma=0.2 $ meV, the parent SC gap closes at $ V_z=1.2 $ meV, the spin-orbit coupling $ \alpha_R=0.5 $ eV\AA, the phenomenological dissipation is $ 3\times 10^{-3} $ meV~\cite{liu2017role}, and zero temperature.}
	\label{fig:3}
\end{figure*}
In Fig.~\ref{fig:3}, we show our calculated conductance spectra corresponding to the experimental results shown in Fig.~\ref{fig:2} using a Gaussian disorder distribution for $ V(x) $ with the disorder parameters shown in the figure caption.  Note that the disorder is spatially fixed once chosen using the Gaussian distribution (for Fig.~\ref{fig:3}).  We adjust the strength of the disorder to get the striking agreement between our theoretical results in Fig.~\ref{fig:3} with the experimental results of Fig.~\ref{fig:2}, but we do not vary the disorder strength once chosen and all the results shown in Fig.~\ref{fig:3} use identical disorder.  There is no disorder averaging here, just one fixed disorder configuration, characterized by the single parameter of disorder strength, since the low experimental temperature implies that ensemble averaging is inappropriate.  The striking agreement between Figs.~\ref{fig:2} and~\ref{fig:3} in both the magnetic field and the tunnel barrier dependence between experiment and theory is remarkable at a qualitative and semiquantitative level.  A quantitative comparison is unfeasible since the experimental disorder parameters are completely unknown and, in fact, experimentally $ V_z $ and $ \mu $ are not known either! (We note that knowing the experimental magnetic field is not equivalent to knowing the Zeeman energy since the precise Land\'e $ g $ factor is not known under the actual experimental conditions~\cite{pan2019curvature}.) So, what we do here with a model disorder potential, obtaining excellent qualitative agreement as a function of several different tuning parameters, is the best one can do.

The most important physical point about Fig.~\ref{fig:3}, and by inference about the experimental results in Fig.~\ref{fig:2}, is that the sharp ZBCPs with large conductance $ \sim 2e^2/h $ are trivial peaks occurring below the TQPT in the trivial $ V_z<V_c $ regime.  Note that in our theory we know the TQPT location $ V_c $ analytically by construction (the bulk gap closes and then reopens at $ V_c $) whereas the TQPT is unknown in the laboratory since a gap reopening is not observed experimentally.  For our parameters in Fig.~\ref{fig:3}, the TQPT is at $ V_c=1.02 $ meV whereas the ZBCPs are at $ V_z \sim 0.92  \text{meV} < V_c$.  In addition, as already emphasized recently by us~\cite{pan2021quantized}, the ZBCP values in Fig.~\ref{fig:3} (as in the experiments shown in Fig.~\ref{fig:2}) are slightly (by about 10\%) larger than the Majorana quantization value of $ 2e^2/h $, further reinforcing the fact that these are not MZM-induced topological ZBCPs, but disorder-induced trivial ZBCPs.  The fact that one can have large (and even sharp) ZBCPs of $ O(2e^2/h) $ arising just from disorder has recently been emphasized~\cite{pan2020generic}, nevertheless, the agreement between Figs.~\ref{fig:2} and~\ref{fig:3} surprised us a great deal, giving us confidence in claiming that it is possible, even likely, that the best Majorana nanowire experiments are disorder-limited in spite of the fact that the induced gap is hard rather than soft as was common in the earlier Majorana experiments~\cite{mourik2012signatures,deng2012anomalous,churchill2013superconductornanowire,das2012zerobias,finck2013anomalous}, where the role of strong disorder was obvious just by virtue of very soft induced gap implying the presence of considerable subgap fermionic states.  The fact that disorder may dominate the ZBCPs even when the zero-field-induced gap is hard (and the ZBCPs themselves are sharp and large) is sobering. There has been earlier theoretical work in the literature on disorder effects in Majorana nanowires, mostly emphasizing class D antilocalization peaks and soft gap arising from the disorder-induced fermionic subgap states as well as on the stability of the Majorana zero modes against weak disorder~\cite{brouwer2011probability,brouwer2011topological,hui2015bulk,liu2012zerobias,lobos2012interplay,sau2013density,bagrets2012class,takei2013soft,cole2015effects,pikulin2012zerovoltage,degottardi2013majorana,adagideli2014effects,sau2012experimental,woods2020enhanced}.  The fact that disorder by itself can, under certain circumstances, give rise to large and sharp trivial zero-bias conductance peaks closely mimicking the Majorana zero-bias peaks, even when the gap is hard, was not realized before. Certainly, no earlier theory presented such a convincing agreement between theory (Fig.~\ref{fig:3}) and experiment (Fig.~\ref{fig:2}) based on trivial disorder-induced theoretical zero-bias conductance peaks.

\begin{figure}[ht]
	\centering
	\includegraphics[width=3.4in]{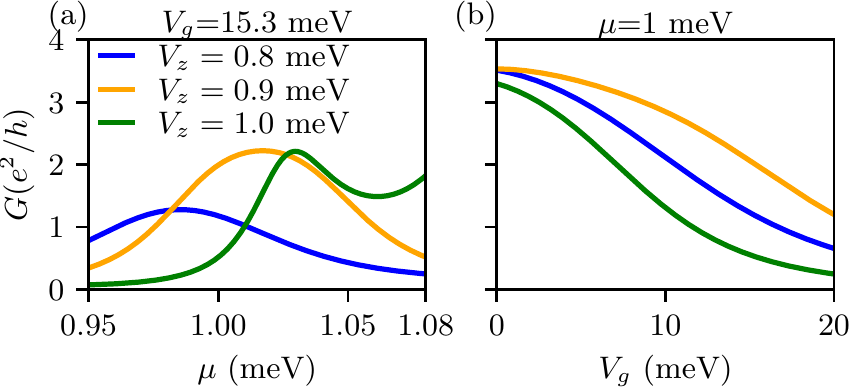}
	\caption{
		(a) The calculated zero-bias tunnel conductance as a function of the chemical potential $ \mu $ at a fixed tunnel barrier $ V_g=15.3 $ meV for $ V_z=$0.8 meV (blue), 0.9 meV (orange), 1 meV  (red);
		(b) The calculated zero-bias tunnel conductance as a function of the tunnel barrier $ V_g $ at a fixed chemical potential $ \mu=1 $ meV for $ V_z=$ 0.8 meV (blue), 0.9 meV (orange), 1 meV  (red). Refer to Fig.~\ref{fig:3} for the rest parameters.}
	\label{fig:4}
\end{figure}

In Fig.~\ref{fig:4}, we further study and discuss the (lack of) stability of the trivial ZBCPs of Fig.~\ref{fig:3} by showing [Fig.~\ref{fig:4}(a)] the calculated zero-bias conductance, at fixed tunnel barrier and Zeeman splitting, as a function of the chemical potential, taking care to ensure that the system stays in the trivial phase with $ V_z<V_c $ always in the whole range of the chemical potential.  We show the calculated chemical-potential-dependent conductance for three values of $ V_z $.  It is clear that the conductance can be tuned almost at will by changing the chemical potential (which is experimentally controlled by various gate voltages, such as the side gate and back gate voltages in Refs.~\onlinecite{zhang2018quantizeda,zhang2021large}).  Although the dependence on the chemical potential is complicated, and nonmonotonic, in Fig.~\ref{fig:4} the key feature is that the calculated conductance can be tuned by tuning chemical potential, with no stability, as also seen experimentally~\cite{zhang2021large}.   In Fig.~\ref{fig:4}(b), we depict the calculated zero-bias conductance as a function of the tunnel barrier potential for fixed Zeeman energy and chemical potential, showing that the trivial ZBCPs of Fig.~\ref{fig:3} vary smoothly with varying tunnel barrier, indicating its trivial character.  The topological MZM manifests a constant $ 2e^2/h $ ZBCP for all tunnel barrier values as long as the temperature is lower than the tunneling energy.  The fact that the ZBCP varies with the tunnel barrier means that it is not quantized, and does not arise from MZMs.  One point to note here is that it is a huge challenge to connect our theoretical parameters $ V_g $ and $ \mu $ to various gate voltages in the experimental situation, and it is likely that a variation in any of the experimental gate potentials, in fact, varies both $ \mu $ and $ V_g $ in some complex manners. For our discussion, however, this is irrelevant because the point we are making is that even when a fine-tuned ZBCP is apparently ``quantized'' (see Fig.~\ref{fig:3}), it is simply a feature of fine tuning and postselection, and the conductance can really be tuned to almost any value between 0 and $ 4e^2/h $ by carefully tuning $ V_z $, $ \mu $, and $ V_g $ in the theory (and, equivalently, by the magnetic field, various gate voltages, and the tunnel barrier in the experiments).

\begin{figure}[t]
	\centering
	\includegraphics[width=3.4in]{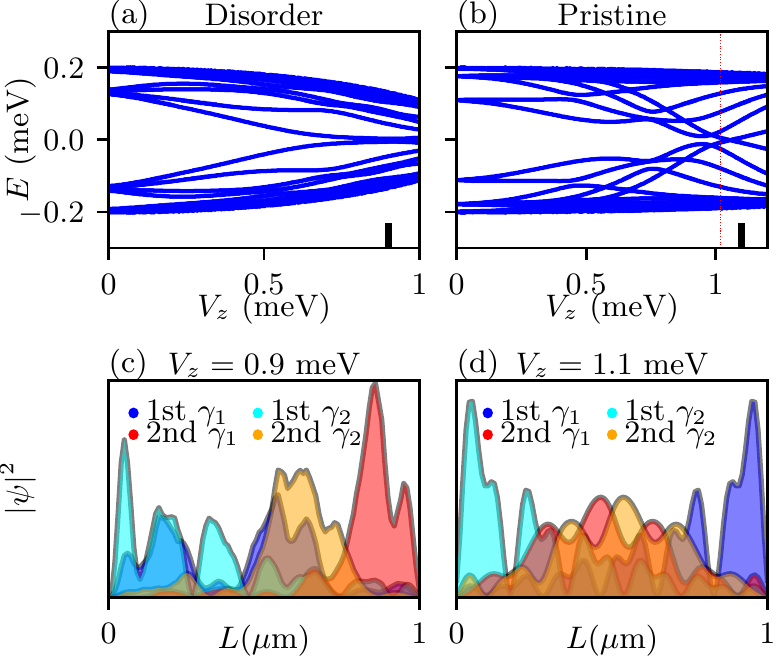}
	\caption{The energy spectra (upper panels) and wave functions (lower panels) of the disordered (left panels) and pristine nanowires (right panels) InSb/Al hybrid nanowire. 
	(a) The energy spectrum as a function of Zeeman field in the presence of disorder, which corresponds to Fig.~\ref{fig:3}.
	(b) The energy spectrum as a function of Zeeman field in a pristine nanowire for the comparison. The topological regime is where $ V_z>1.02 $ meV indicated by the vertical red dashed line. The SC gap always persists as Zeeman field increases.
	(c) shows the lowest and second-lowest wave functions in the trivial regime at $ V_z=0.9 $ meV corresponding to the black line in (a);
	(d) shows the lowest and second-lowest wave functions in the topological regime at $ V_z=1.1 $  meV corresponding to the black line in (b). Refer to Fig.~\ref{fig:3} for the other parameters (except for the tunnel barrier and dissipation, which are absent here).}
	\label{fig:5}
\end{figure}
In Fig.~\ref{fig:5} we show the calculated energy spectra (and the lowest {and the second-lowest} wave functions) with and without the disorder potential to emphasize what is going on at the microscopic quantum-mechanical level. {Figure~\ref{fig:5}(a) shows the energy spectrum in the presence of disorder corresponding to Figs.~\ref{fig:3}(a)-\ref{fig:3}(c) (since they only differ by the height of the tunnel barrier). The pair of low-lying states emerging from $ V_z=0.75 $ to 1 meV is trivial because the putative TQPT is at $ 1.02 $ meV. Given the finite dissipation (or even low, but finite, temperature), this pair of low-lying states may merge into a single peak and thus manifest the zero-bias peaks in Fig.~\ref{fig:3}. We also plot the corresponding wavefunction in the Majorana basis at $ V_z=0.9 $ meV below the TQPT in Fig.~\ref{fig:5}(c). The wave functions of the lowest state are in blue and cyan while the wave functions of the second-lowest state are in red and orange. We find that the two Majorana modes are highly overlapping, which indicates a trivial fermionic state. For comparison, we also present the energy spectrum of the corresponding disorder-free pristine nanowire in Fig.~\ref{fig:5}(b). The TQPT happens at $ V_z=1.02 $ meV (red dashed line); thus, the corresponding wavefunction at $ V_z=1.1 $ meV [Fig.~\ref{fig:5}(d)] is in the topological regime. The wave function of the lowest state is localized at two ends of the nanowire, while the wavefunction of the second-lowest state is localized in the bulk region of the nanowire.}

Several aspects of our disordered Majorana nanowire results need to be emphasized in order to avoid any misunderstanding.  First, there is no ensemble averaging, the quenched spatial disorder is one fixed random configuration chosen from a Gaussian distribution, consistent with the experiment being done at very low temperatures.  Second, most disorder configurations with the same strength and variance do not produce large zero-bias trivial peaks, in fact, most disordered results have small zero-bias peaks or no peaks at all, only a few configurations give rise to large zero-bias peaks.  This is, of course, completely consistent with the experimental situation where the protocol is to go through many samples varying tuning parameters until large zero-bias peaks appear in the measurements, and once such a peak appears, it is fine tuned to produce the desired results.  Both the experiment~\cite{zhang2021large} and our theory produce only of the order of $ \sim $2\% samples manifesting large zero-bias peaks with conductance  $ \sim O(2e^2/h) $.  Large zero-bias peaks are not generic by any means either in the experiment or in the theory with disorder, only for the pristine nanowires with real topological Majorana modes, zero-bias peaks with conductance $ 2e^2/ h $ appear generically for $ V_z $ above the TQPT. We mention as an aside that very stable trivial ZBCPs with values pinned at $ 2e^2/h $  may arise from certain types of smooth deterministic potential-induced trivial tunneling peaks, but such stable peaks do not go above $ 2e^2/h $; we have recently studied the trivial ZBCPs arising from smooth potentials in great depth in a recent work~\cite{pan2021quantized}; (see also~\cite{pan2020physical,moore2018twoterminal,kells2012nearzeroenergy,fleckenstein2018decaying,prada2012transport,liu2018distinguishing,moore2018quantized,vuik2019reproducing,reeg2018zeroenergy}). {We believe that the measurements of Refs.~\onlinecite{zhang2018quantizeda,zhang2021large,yu2021nonmajorana}, as well as all the earlier nanowire Majorana experiments, are dominated by random disorder effects and not by any smooth potential-induced quasi-Majorana behavior.} The other thing to emphasize is that even our disordered system eventually would manifest topological zero-bias peaks at large enough Zeeman field well above the TQPT so that the Zeeman field can overcome the disorder effect.  But, experimentally such a high-field regime is currently inaccessible since the bulk superconductivity is completely suppressed at high field in the experiment, most likely because the parent Al superconductivity is quenched by the high magnetic field.  The zero-bias peaks in Fig.~\ref{fig:3} in our theory are all nontopological occurring below TQPT, and are induced entirely (but only occasionally, not generically) by disorder. No experiment has ever reported either ZBCPs  $\sim2e^2/h $ values stable over a large magnetic field range or reentrant ZBCPs with $ \sim 2e^2/h $ conductance at high magnetic fields, so we can safely conclude that the existing experimental data do not provide evidence for any topological ZBCP at $ V_z>V_c $.   We emphasize that the good MZM-induced topological ZBCPs could manifest conductance $ < 2e^2/h $ (e.g., because of temperature effect), but not $ >2e^2/h $, as in Fig.~\ref{fig:2} and Ref.~\onlinecite{zhang2021large}.

\begin{figure}[t]
	\centering
	\includegraphics[width=3.4in]{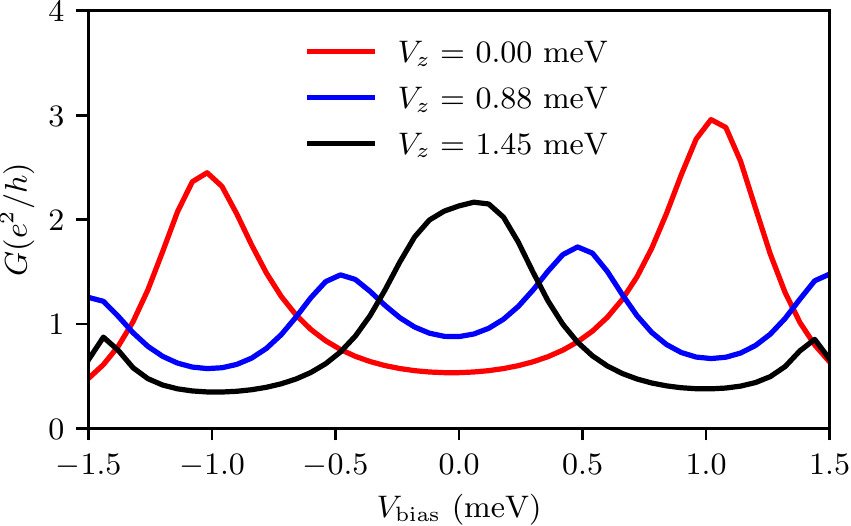}
	\caption{The tunneling conductances in the trivial regime as a function of bias voltages at fixed Zeeman fields 0 meV (red), 0.88 meV (blue), and 1.45 meV (black) to reproduce Fig. 3(c) in Ref.~\onlinecite{yu2021nonmajorana} using exactly the same Hamiltonian as in Fig.~\ref{fig:3} but only with a different set of SC parameters corresponding NbTiN: the parent SC gap is 3 meV and the SC-SM coupling strength is 1.5 meV (hence the induced SC gap is around 1 meV)~\cite{lutchyn2018majorana}. The other parameters are as follows: the wire length is 0.4 $\mu $m, the chemical potential is $ \mu=5 $ meV, the variance of disorder $ \sigma_\mu=20 $ meV, parent SC gap closes at $ V_z=10 $ meV, the spin-orbit coupling $ \alpha_R=0.5 $ eV\AA, the phenomenological dissipation is 0.1 meV, and the tunnel barrier height $ V_g $ is 10 meV.  We also verify that the conductance on the other end [left end in Fig.~\ref{fig:8}(c)] of the nanowire has no feature, namely, no subgap states emerge and the conductances are almost zero everywhere inside the parent SC gap.}
	\label{fig:6}
\end{figure}

Now, we briefly discuss a very recent experiment, where very small ZBCPs $ \sim 2e^2/h $ are reported in InSb/NbTiN SM/SC structures, with the system being well-known to be very highly disordered~\cite{yu2021nonmajorana}.  In fact, very similar (essentially identical) tunneling data to what is reported in~\cite{yu2021nonmajorana}  were already reported extensively in the early experiments on Majorana nanowires~\cite{mourik2012signatures,deng2012anomalous,churchill2013superconductornanowire,das2012zerobias,finck2013anomalous}, where the characteristic features were the existence of a soft SC gap (indicating huge disorder in the system) and very non-sharp small peaks over the large background tunnel conductance because of the presence of considerable subgap fermionic states (the same states leading to the gap being soft).  These are the same disorder-induced features showing up prominently in the new work in 2021~\cite{yu2021nonmajorana}. There is nothing new in the data of~\cite{yu2021nonmajorana} except for the emphasis on the ZBCPs being of $ O(2e^2/h) $ which was true for several early experiments too, although not specifically emphasized in those publications.  In addition, the new work emphasizes the fact that tunneling from only one end manifests ZBCPs, not from both ends, further reinforcing the trivial nature of these small (above the background) ZBCPs.  In Fig.~\ref{fig:6}, we show our fine-tuned simulated theoretical results corresponding to the observations of~\cite{yu2021nonmajorana} using nanowire parameters corresponding to InSb/NbTiN system, where the parent SC gap is larger (than in Al) and so are disorder and dissipation (arising from the copious presence of vortices in the rather poor quality NbTiN material). It is well-known that NbTiN is a very poor quality SC which is why nobody uses it in Majorana experiments any longer, and the rationale for using such a highly disordered parent SC in Ref.~\onlinecite{yu2021nonmajorana} defies logic in 2021.  We reproduce disorder-induced small ZBCPs above the background which are not sharp at all (similar to the experiment in~\cite{yu2021nonmajorana}), but have conductance values $ \sim 2e^2/h $.  Our results of Fig.~\ref{fig:6} look very similar to the results shown in Figs. 2(c) and 3(c) of~\cite{yu2021nonmajorana}.  We note that neither experiment~\cite{yu2021nonmajorana}  nor our theory generically produces ZBCPs $ \sim O(2e^2/h) $ --- both are fine-tuned results without much significance or import. The characteristic features are as follows:  the ZBCPs are trivial arising entirely from disorder, a strong asymmetry in the peaks arising from dissipation~\cite{liu2017role}, very soft SC gap because of disorder-induced subgap states, and small non-sharp ZBCPs above the background with conductance $ \sim O(2e^2/h) $.  We have checked that the tunneling from the other end here does not produce any ZBCPs in our simulations, which is of course expected, and we have also checked that these calculated ZBCPs are not robust at all against the tunnel barrier potential. In fact, the tunnel barrier potential has to be carefully tuned both in our theory and in the experiment of~\cite{yu2021nonmajorana} in order to find small peaks with $ 2e^2/h $ values: it is a result of precise fine tuning in magnetic field, tunnel barrier, and disorder potential.  The peaks in the experiment of~\cite{yu2021nonmajorana} and in our Fig.~\ref{fig:6} are similar to the ones seen in the early 2012 nanowire tunneling spectroscopies~\cite{mourik2012signatures,deng2012anomalous,churchill2013superconductornanowire,das2012zerobias,finck2013anomalous}, and are by no means similar to the results in Figs.~\ref{fig:2} and~\ref{fig:3}, i.e., in Ref.~\onlinecite{zhang2021large}, where the gap is hard and the ZBCPs are large and sharp.  We do not believe that the experimental observations of Refs.~\onlinecite{zhang2021large,yu2021nonmajorana} are equivalent as the latter is reporting results in extremely highly disordered systems with a huge amount of subgap fermionic states as was the situation in 2012.  It is clear that sample improvement during 2012--2021 has vastly improved the quality of the data in Majorana nanowires, and we believe that further improvement in sample quality will lead to the observation of topological Majorana modes. 
 
\begin{figure}[t]
	\centering
	\includegraphics[width=3.4in]{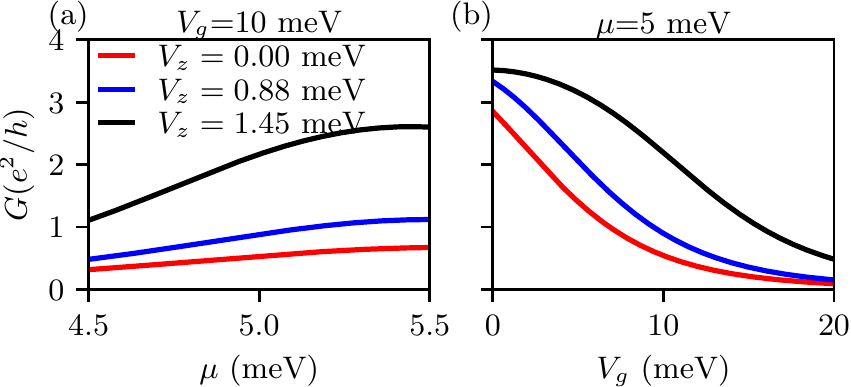}
	\caption{
	(a) The calculated zero-bias tunnel conductance as a function of the chemical potential $ \mu $ at a fixed tunnel barrier $ V_g=10 $ meV for $ V_z=$0 meV (red), 0.88 meV (blue), and 1.45 meV (black);
	(b) The calculated zero-bias tunnel conductance as a function of the tunnel barrier $ V_g $ at a fixed chemical potential $ \mu=5 $ meV for $ V_z=$0 meV (red), 0.88 meV (blue), 1.45 meV (black). Refer to Fig.~\ref{fig:6} for the rest parameters.
	}
	\label{fig:7}
\end{figure}

\begin{figure}[t]
	\centering
	\includegraphics[width=3.4in]{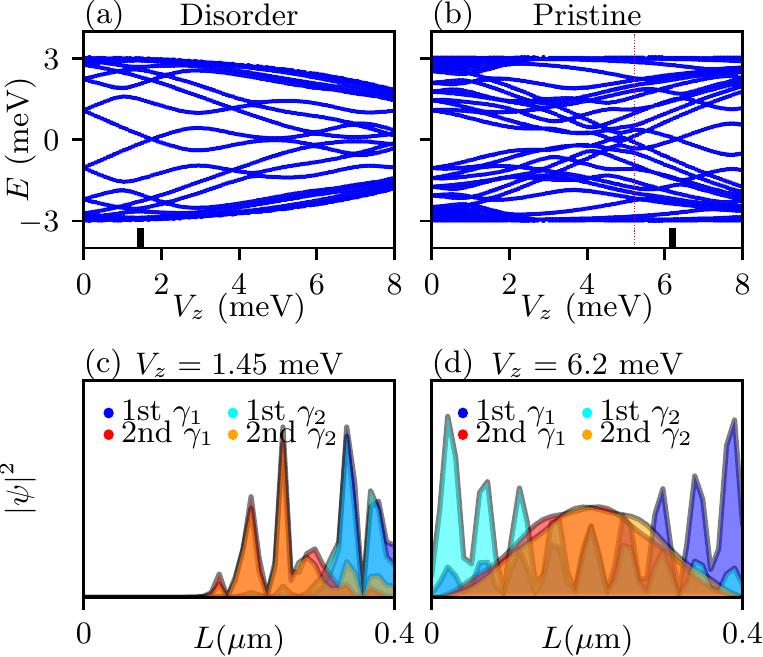}
	\caption{The energy spectra (upper panels) and wave functions (lower panels) of the disordered (left panels) and pristine nanowires (right panels) in InSb/NbTiN hybrid nanowire. 
	(a) The energy spectrum as a function of Zeeman field in the presence of disorder, which corresponds to Fig.~\ref{fig:6}.
	(b) The energy spectrum as a function of Zeeman field in a pristine nanowire for the comparison. The topological regime is where $ V_z>5.22 $ meV indicated by the vertical red dashed line. The SC gap always persists as Zeeman field increases;
	(c) shows the lowest and second-lowest wavefunctions in the trivial regime at $ V_z=1.45 $ meV corresponding to the black line in (a);
	(d) shows the lowest and second-lowest wavefunctions in the topological regime at $ V_z=6.2 $  meV corresponding to the black line in (b). Refer to Fig.~\ref{fig:6} for the other parameters (except for the tunnel barrier and dissipation, which are absent here). }
	\label{fig:8}
\end{figure}
Another key aspect of Ref.~\onlinecite{yu2021nonmajorana}, not discussed much in the paper, is that the authors had to do an arbitrary contact resistance subtraction in order to make their ZBCPs $\sim O(2e^2/h) $, and the subtracted resistance itself is of $ O(h/3e^2) $, implying that the unsubtracted ZBCPs in~\cite{yu2021nonmajorana} have values much less than $ 2e^2/h $!  Such an arbitrary resistance subtraction is essentially confirmation bias in its fullest glory; the authors wanted to prove that their tiny peaks are $ \sim O(2e^2/h) $, and they did so by suitably subtracting arbitrary resistance from their measured tunnel conductance.  We give no credence nor significance to the results in~\cite{yu2021nonmajorana}.  While it might be possible to unwittingly suggest (incorrectly of course, as originally happened in Refs.~\onlinecite{zhang2018quantizeda,zhang2021retraction}) the high-quality conductance results of Ref.~\onlinecite{zhang2021large} shown in Fig.~\ref{fig:2} as the manifestation of MZM quantization erroneously, we simply do not see anybody ever mistaking, under any circumstances, the very poor-quality small peaks of~\cite{yu2021nonmajorana} as MZM quantization ever (even without taking into account the problem with the arbitrary resistance subtraction), and we do not, therefore, understand at all what misleading point is being made in~\cite{yu2021nonmajorana} by claiming the appearance of $ 2e^2/h $ conductance peaks.

In Fig.~\ref{fig:7} (see Fig.~\ref{fig:4} for comparison with the same results for Ref.~\onlinecite{zhang2021large} parameters), we show our calculated conductance for Ref.~\onlinecite{yu2021nonmajorana} parameters as a function of chemical potential and tunnel barrier keeping all other quantities fixed (and ensuring that the system is always trivial, i.e., $ V_z < V_c $ throughout).  As expected, the conductance at zero-bias varies smoothly as a function of either the chemical potential [Fig.~\ref{fig:7}(a)] or the tunnel barrier [Fig.~\ref{fig:7}(b)], reflecting their trivial unstable character, similar to what we find in Fig.~\ref{fig:4} for the Delft sample.  Thus, trivial ZBCP, arising at $ V_z < V_c $, could possibly be tuned to $ \sim 2e^2/h $ (or to any value between 0 and $ 4e^2/h $) by fine tuning the chemical potential and/or the tunnel barrier strength, but such a fine-tuned $ 2e^2/h $ ZBCP is neither stable nor meaningful.  Of course, for the extremely disordered samples used in Ref.~\onlinecite{yu2021nonmajorana}, these peaks are very small and are additionally misleadingly fine-tuned by carrying out an arbitrary contact resistance subtraction, making the whole fine tuning exercise meaningless and bizarre. It appears that in their zeal to establish that the reported ZBCP quantization in Ref.~\onlinecite{zhang2021retraction} is unstable and not associated with topological MZMs, something the two of the current authors already pointed out very clearly using theoretical arguments and detailed simulations some time ago~\cite{pan2020physical}, Ref.~\onlinecite{yu2021nonmajorana} has actually done a rather futile analysis involving arbitrary resistance subtractions for highly resistive small conductance peaks.  We feel that Ref.~\onlinecite{yu2021nonmajorana} establishes nothing other than the fact that one can usually get anything one wants by arbitrarily subtracting one large number from another large number.  

In Fig.~\ref{fig:8} (c.f., Fig.~\ref{fig:5} for the corresponding results for the corresponding Delft~\cite{zhang2021large} results) we show our calculated energy spectra and low-lying wave functions for the typical parameters of Ref.~\onlinecite{yu2021nonmajorana}.  The important point of Fig.~\ref{fig:8} is that disorder mixes the Majorana states from the two ends strongly, and depending on the details of this overlap, it is possible to have a trivial zero-bias tunneling peak from one end, but not from the other end.

\section{Discussion and Conclusion}\label{sec:III}

We have provided in this work detailed Majorana nanowire theoretical simulations, showing that trivial zero-bias conductance peaks with sharp and large $ \sim O(2e^2/h) $ values may appear in the tunneling spectra mimicking aspects of topological Majorana zero modes, thus misleading the experimentalists.  These large trivial peaks arise from random disorder with fine tuning very similar to the experimental protocol used in the experimental data analysis searching for Majorana zero modes through tunneling spectroscopy.  Just as not all laboratory samples manifest large zero-bias peaks, not all random disorder configurations produce large trivial peaks, and therefore the possibility of confirmation bias is considerable because the inevitable presence of disorder in the experimental samples may very well produce large zero-bias tunnel conductance peaks through extensive data selection, which is the current laboratory protocol for the Majorana search.  For example, Zhang \textit{et al.}~\cite{zhang2021large,zhang2018quantizeda} find the reported large zero-bias peaks, which they originally mistakenly attributed as ``Majorana quantization'', only in 2 out of around 80 samples they looked at!  This is a $ \sim $2\% yield on the desired outcome one is trying to find, and should always be taken with considerable suspicion as possibly being ``false positives'' arising from the inherent nature of the confirmation bias syndrome.  We believe that it is likely that all existing claims of Majorana observations in the literature are trivial peaks arising from disorder, but of course we can only suggest this as a real possibility, we cannot prove it. Something negative, i.e., true MZMs have not yet been experimentally observed, can only be persuasively suggested by our theory, by definition, it cannot be proven. The fact that the observed zero-bias peaks never manifest any robust stability in experimental tuning parameters, e.g., the magnetic field, gate voltage, and tunnel voltage, which is consistent with our disorder-induced trivial peaks in this work, where we find that the disorder-induced trivial ZBCPs show smooth variations in Zeeman energy, chemical potential, and barrier potential, and inconsistent with topological zero modes also lends credence to our suggestion that Majorana experiments are dominated by random disorder.  We would speculate that this disorder problem actually is prevalent in most, if not all, topological experiments in the literature, and not just  Majorana experiments: we must remember that no robust and stable topological quantization has ever been reported in any experimental system other than quantum Hall experiments. Although our current work focuses on SC-SM nanowire platforms, because the best and the most quantitatively compelling tunneling spectroscopy data are extensively available in Majorana nanowires, other systems, where the zero-bias peaks are studied at vortex cores using scanning tunneling microscopy, manifest zero-bias peaks of extreme small values which are much more consistent with disorder-induced subgap Andreev states in the vortex cores than with topological MZMs.

Perhaps the most important finding in this work is its remarkable agreement with the reported data in Ref.~\onlinecite{zhang2021large}: both experiment~\cite{zhang2021large} and our theory find similar-looking large and sharp fine-tuned zero-bias tunnel conductance peaks which are not robust against variations in Zeeman field, chemical potential, and tunnel barrier, and both manifest peaks which could go above $ 2e^2/h $ quantization value, but may be fine tuned (by adjusting parameters such as magnetic field and gate voltages) to $ 2e^2/h $, enhancing the unfortunate possibility of a confirmation-bias-induced claim for Majorana quantization. Also, our trivial ZBCPs generically exist only for tunneling from one end, and not the other end, although that can occasionally happen accidentally~\cite{pan2020physical,lai2019presence}. Therefore, we urge experimentalists to try simultaneous tunnel spectroscopies from both ends as an additional distinguishing aspect of trivial versus topological.
 
{We note that the important role of disorder in producing trivial zero-bias conductance peaks in Majorana nanowires has earlier been considered in the literature (see, e.g., Ref.~\onlinecite{pan2020physical} and references therein).  In this work, we apply the standard disorder theory to understand in depth the recent experiments reported in Refs.~\onlinecite{zhang2018quantizeda,zhang2021large,yu2021nonmajorana}, establishing convincingly, through detailed comparison with the experimental data, that the large zero-bias conductance peaks reported in these experiments are likely to be disorder-induced trivial peaks, which could accidentally achieve conductance values $ \sim O(2e^2/h) $.  Such trivial disorder-induced peaks are generic in class D systems~\cite{pan2020generic}, and are inevitably present if the results are postselected from a large amount of data.  Our results clearly show that seeing occasional large zero-bias conductance peaks is not evidence for the existence of topological Majorana modes.  Our excellent agreement with the experimental observations indicates the need for producing much higher-quality nanowire samples with much less disorder for future Majorana experiments.}

Our work suggests that serious vigilance is necessary to guard against claims of topological discoveries based on confirmation bias arising from precise existing theoretical predictions.  In particular, all such experimental claims should necessarily release all data collected in the experiment (including data which are inconsistent with the topological predictions) so that the community could go through the data to ensure that the outcome is not generated by confirmation bias achieved through fine-tuned data selection. The problem that arose in Refs.~\onlinecite{zhang2018quantizeda,zhang2021large} could of course always happen in spite of one's best efforts, and the best scientific practice would then be an immediate retraction if the subsequent analysis indicates that the original claim of discovery is an unwitting false-positive finding. The fundamental problem here is that, given sufficient number of tunable parameters (magnetic field, tunnel barrier, gate voltage) in the experiment, it is often, if not always, possible to keep on tuning parameters until one finds precisely what one is looking for.  What we show here is that disorder produces trivial peaks, which on sufficient fine tuning, would produce $ \sim 2e^2/h $ peaks, and this by itself is no discovery, it is simply confirmation bias.

One specific and concrete conclusion of our extensive theoretical work on Majorana nanowires including effects of disorder is that topological Majorana zero modes in all likelihood have not yet been observed in the laboratory since even the very best currently available experimental data (of Ref.~\onlinecite{zhang2021large}) appear consistent with disorder-induced zero-bias peaks in the theory.  There has been enormous progress since 2012, with the emergence of hard zero-field gaps with no obvious features of deleterious subgap fermionic states and very large and sharp fine-tuned zero-bias peaks at finite Zeeman splitting, but we still need purer samples with less disorder to observe truly topological Majorana zero modes at finite magnetic fields.  We suggest materials improvement as the most essential necessity in this field. The other key general conclusion of our work is that confirmation bias is almost inevitable in claiming topological discoveries if one looks for something precise among a huge amount of data, particularly if there are many parameters to tune and many samples to use.  Let this be a cautionary tale for the whole topological condensed matter physics since this field is currently all about experiments chasing precisely known theoretical predictions.

This work is supported by the Laboratory for Physical Sciences. This research is also supported in part by the Heising-Simons Foundation, the Simons Foundation, and National Science Foundation Grant No. NSF PHY-1748958. We acknowledge the University of Maryland High-Performance Computing Cluster (HPCC). 

\bibliography{Paper_disorder}
\end{document}